\documentclass[11pt,preprint]{aastex}

\usepackage{verbatim}
\usepackage{natbib}
\usepackage{amsmath}
\usepackage{amsbsy}
\usepackage{lscape}

\shorttitle{Turbulent high-redshift galaxies}
\shortauthors{Burkert, Genzel, Cresci, Bouche, Naab et al.}

\begin{document}

\title{High-Redshift Star-Forming Galaxies: Angular Momentum and Baryon Fraction, Turbulent Pressure Effects and the Origin of Turbulence}

\author{A. Burkert\altaffilmark{1,2}, R. Genzel\altaffilmark{3}, N. Bouch\'{e}\altaffilmark{3}, G. Cresci\altaffilmark{3},
S. Khochfar\altaffilmark{3}, J. Sommer-Larsen\altaffilmark{4,5}, A. Sternberg\altaffilmark{6}, T. Naab\altaffilmark{1}, N. F\"orster Schreiber\altaffilmark{3},
L. Tacconi\altaffilmark{3}, K. Shapiro\altaffilmark{7}, E. Hicks\altaffilmark{3}, D. Lutz\altaffilmark{3}, R. Davies\altaffilmark{3}, P. Buschkamp\altaffilmark{3},
S. Genel\altaffilmark{3}}

\altaffiltext{1}{University Observatory Munich (USM), Scheinerstrasse 1, 81679 Munich, 
Germany}
\altaffiltext{2}{Max-Planck-Fellow}
\altaffiltext{3}{Max-Planck-Institut f\"ur extraterrestrische Physik (MPE), Giessenbachstr. 1,
85748 Garching, Germany}
\altaffiltext{4}{Dark Cosmology Centre, Niels Bohr Institute, University of Copenhagen, Juliane Marie
Vej 30, 2100 Copenhagen, Denmark}
\altaffiltext{5}{Excellence Cluster Universe, Technical University Munich, Boltzmannstr. 2, D-85748 Garching, 
Germany}
\altaffiltext{6}{School of Physics and Astronomy, Tel Aviv University, Tel Aviv 69978, Israel }
\altaffiltext{7}{Department of Astronomy, Campbell Hall, University of California, Berkeley, CA 94720, USA }

\email{burkert@usm.uni-muenchen.de, genzel@mpe.mpg.de}

\begin{abstract}

The structure of a sample of high-redshift ($z \sim 2$), rotating 
galaxies with high star formation rates and turbulent gas velocities of
$\sigma \approx 40-80$ km/s is investigated. Fitting the observed
disk rotational velocities and radii with a \cite{mo98}
(MMW) model requires unusually large disk
spin parameters $\lambda_d > 0.1$ and disk-to-dark halo mass fractions
of m$_d \approx 0.2$, close to the cosmic baryon fraction. The galaxies segregate
into dispersion-dominated systems with $1 \leq v_{max}/\sigma \leq 3$, maximum rotational velocities
$v_{max} \leq $ 200 km/s and disk half-light radii $r_{1/2} \approx$ 1-3 kpc
and rotation-dominated systems with $v_{max} > $ 200 km/s,
$v_{max}/\sigma > 3$ and $r_{1/2} \approx$ 4-8 kpc. For the dispersion-dominated sample 
radial pressure gradients
partly compensate the gravitational force, reducing the rotational velocities. 
Including this pressure effect in the MMW model, dispersion-dominated
galaxies 
can be fitted well with spin parameters of $\lambda_d = 0.03-0.05$ for high disk
mass fractions of m$_d \approx 0.2$ and with $\lambda_d = 0.01-0.03$ for m$_d \approx 0.05$.
These values are in good agreement with cosmological expectations.
For the rotation-dominated sample however pressure effects are small and
better agreement with theoretically expected disk spin parameters can only be achieved if the 
dark halo mass contribution in the visible disk regime
($2-3 \times r_{1/2}$) is  smaller than predicted by the MMW model. We argue that
these galaxies can still be embedded in standard cold dark matter halos if the 
halos did not contract adiabatically in response to disk formation. In this
case, the data favors models with small disk mass fractions of m$_d = 0.05$ and
disk spin parameters of $\lambda_d \approx 0.035$.
It is shown that the observed high turbulent gas motions of the galaxies are
consistent with a Toomre instability
parameter Q=1 which is equal to the critical value, expected for
gravitational disk instability to be the major driver of turbulence. 
The dominant energy source of turbulence is then the potential energy of the gas in the disk.
\end{abstract}


\section{Introduction}

Deep surveys have become efficient in detecting star-forming galaxy populations 
at z $\sim$ 1.5-3.5, near the peak of cosmic star formation, the assembly of massive galaxies 
and QSO activity (e.g. \citep{ste96,ste04,fra03,dad04b}). 
Large samples are now available, based on their rest-frame UV, or optical, magnitude/color properties. 
These high-redshift galaxies have star formation rates of 10-300 M$_{\odot}$/yr, with a range of ages 
(10 Myrs - 3 Gyrs), stellar masses of M$_* \sim 10^9 - 10^{11.5}$ M$_{\odot}$ \citep{sha05,foe06,
erb06a,erb06b,dad04a,dad04b} and high gas fractions \citep{tac10}. 
They contribute a large fraction of the cosmic star formation activity and stellar mass density at z $\sim$ 2 
\citep{red05,rud06,van06,gra07,per08}. The majority of these galaxies appears to form stars with 
high rates over a significant fraction of the z $\sim$ 1.5-3 redshift range \citep{dad07,noe07}.
This requires an efficient and semi-continuous replenishment of fresh gas, perhaps delivered by cold 
flows/streams from the halo \citep{dek06,dek09a,dek09b,ker05,ocv08,gen08,elm10,aum10}.

High resolution near-infrared integral field spectroscopy of H$\alpha$ line emission has shown that most of 
these high-z star forming galaxies are clumpy and exhibit large ionized gas velocity dispersions of 30-120 km/s 
\citep{foe06,foe09,gen06,gen08,law07,law09,wri07,wri09,van08,sta08,bou08,epi09,cre09}. About one third 
appear to be rotating disks, one third are dispersion dominated systems and one third show clear evidence 
for interactions and major mergers \citep{sha08,foe09}. The fraction of large, 
clumpy rotating disks increases with mass. The ratio of the rotational to random velocities 
ranges between 1 and 6, quite in contrast to z $\sim$ 0 disk galaxies where v/$\sigma \sim$ 10-20 \citep{dib06}. 
Many high-z disks are turbulent and geometrically thick (e.g. \cite{elm06}). 
Important questions are 
what drives and maintains these large turbulent velocities and how turbulence is connected
to the clumpy disk substructure and the high star formation rates \citep{imm04a,imm04b,bou07,bou08,dek09a,dek09b,kho09}.

In addition to the unusual kinematics and structure of high-redshift disks, their
global physical parameters appear to be puzzling and inconsistent with theoretical expectations. \cite{bou07}
found that many high-z galaxies lie in a similar part of the rotation velocity
versus disk radius plane as late-type z $\sim$ 0 disks \citep{cou07} which is not expected according to the
standard \cite{mo98} (MMW) model of galactic disk structure. \cite{foe09} compared the derived dynamical masses with
the stellar masses (from spectral energy distribution analysis) and gas masses (from an application of 
the Kennicutt-Schmidt star formation relation). They found disk masses that are remarkably and perhaps implausibly high.

The MMW model neglects galactic disk turbulence which is
reasonable for present day disks with v/$\sigma \sim$ 20. The situation is however different at
high redshifts where turbulence can strongly affect the disk structure.
This paper discusses the impact of large turbulent motions 
on the interpretation of the dynamical data of disk galaxies. We show that, including
turbulent pressure, the disk spin parameters and disk mass fractions of dispersion-dominated
galaxies are reduced to values that are consistent with 
theoretical expectations. The situation is different for rotation-dominated galaxies
where pressure effects play a minor role.  As already suggested by numerous studies at low redshifts
(e.g. \cite{mo00,dut07}), we
argue that the observed high-redshift galaxies are in better
accord with cosmological models if it is assumed that their dark-matter
halos did not contract adiabatically.  Finally we propose an explanation for why large turbulence might
be more common in many high-z disks and what the energy source of turbulence in these disks might be.

\section{Rotation Curves of Pressurized, Turbulent Galactic Disks}

Let us consider a turbulent galactic gas disk. We analyse its rotational velocity $v_{rot}$ 
in the midplane, applying the hydrostatic equation

\begin{equation}
\frac{v_{rot}^2}{r} = f_g(r) + \frac{1}{\rho} \frac{dp}{dr} 
\end{equation}

\noindent where $r$ is the distance from the galactic center and $f_g$ is the 
value of the gravitational force. $p$ is the pressure which consists of a
turbulent (kinetic) and thermal part, $p = \rho (\sigma^2 + c_s^2)$
with $\rho$ the gas density, $\sigma$ the characteristic 1-dimensional velocity
dispersion of the gas which we assume to be isotropic and $c_s$ its sound speed. We define the zero-pressure
velocity  curve $v_0(r)$ as the rotational velocity of the gas if
pressure gradients are negligible, i.e. dp/dr=0: $v_0^2 \equiv f_g \times r$.
Equation 1 then reduces to 

\begin{equation}
v_{rot}^2=v_0^2+\frac{r}{\rho} \frac{dp}{dr}=v_0^2+\frac{1}{\rho} \frac{d}{d ln r} \left( \rho \sigma^2 \right) .
\end{equation}

\noindent Here we have neglected the thermal pressure term as the sound speed is in general
much smaller than the turbulent velocity. Equation 2 is the most general form, without specifying the radial dependence
of $\sigma$ and $\rho$. It demonstrates that a negative radial pressure gradient
reduces the rotational velocity of the gas as part of the gravitational force
is balanced by the pressure force.  

To illustrate the possible importance of pressure effects,
let us now assume that $\sigma$  is independent of $r$. Then
   
\begin{equation}
v_{rot}^2 = v_0^2+\sigma^2 \frac{d ln \rho}{d ln r} .
\end{equation}

\noindent If
$\sigma$ is also independent of height $z$ above the disk's equatorial plane, the
vertical density distribution is given by the vertical hydrostatic Spitzer solution
(Spitzer 1942; Binney \& Tremaine 08, page chapter 4, p. 390)

\begin{equation}
\rho (z) = \rho_0 sech^2 (z/h)
\end{equation}

\noindent with $\rho_0$(r) the density in the midplane (z=0) at radius r and

\begin{equation}
h=\frac{\sigma}{\sqrt{2 \pi G \rho_0}}
\end{equation}

\noindent the scale height. The total mass surface density $\Sigma(r)$ of such a disk is
(Binney \& Tremaine 2008) 

\begin{equation}
\Sigma =2 \rho_0 h
\end{equation}

\noindent so that

\begin{equation}
\rho_0 = \frac{\pi G \Sigma^2}{2 \sigma^2}
\end{equation}

\noindent The equations 5-7 offer an interesting future observational test
of our assumptions as for an exponential disk (equation 10) with constant velocity
dispersion the scale height h should increase exponentially with radius

\begin{equation}
h=\frac{\sigma^2}{\pi G \Sigma_0} \exp \left( \frac{r}{r_d} \right)
\end{equation}

\noindent Inserting $\rho_0$ from equation 7 into equation 3, the
rotation curve in the equatorial plane of a pressurized gas disk is 

\begin{equation}
v_{rot}^2 = v_0^2 + 2 \sigma^2 \frac{d ln \Sigma}{d ln r}
\end{equation}

\noindent For example, for an exponential disk profile with scale length
$r_d$ 

\begin{equation}
\Sigma (r) = \Sigma_0 \times \exp \left(- \frac{r}{r_d} \right)
\end{equation}

\noindent Equation 9 leads to

\begin{equation}
v_{rot}^2=v_0^2-2 \sigma^2 \left(\frac{r}{r_d} \right)
\end{equation}

\noindent Note that v$_{rot}$ is the actually observable rotational velocity
of the gas, while v$_0$ is the rotation expected if pressure effects 
are negligible ($\sigma = 0$). For $v_{rot}/\sigma \lesssim$ 3 the rotational velocity is
significantly reduced by turbulent pressure effects for $r \gtrsim r_d$.
A very similar equation holds for stellar disks as a special form of the Jeans equation
(Binney \& Tremaine 2008).
Equation 11 was derived for an exponential surface density distribution and
a constant velocity dispersion.  In general, the situation is more complex 
as parts of the disk might have constant surface densities while other parts might 
show a steep gradient. In addition, the velocity dispersion might
change with radius. In this case one would have to solve equation (1) directly.
The best-resolved high-redshift disk galaxies show roughly
constant velocity dispersion profiles (Genzel et al. 2008) and exponentially decreasing
H$\alpha$-surface brightness distributions of the star-forming gas with
scale lengths similar to the stellar disks within 1-3 disk scale lengths (Cresci et al. 2009;
Bouch\'{e} et al., in preparation). 
This is also the region where the rotation curves are flat and achieve their
maximum values $v_{max}$ (Fig. 3). $v_{max}$ will be used in the following
to compare the observations with theory.

For the purpose of this analyses we adopt
equation 11 in order to calculate the pressure corrected rotation curves. For simplicity we will
also assume that gas and stars have similar disk scale lengths, equal to the sizes as 
derived for the star-forming gas from the H$\alpha$ measurements. Note however that there are theoretical reasons
why the scale radius of the gaseous disk component, including the part that is not forming stars
violently, should be larger than the scale length of the stellar disk (e.g. Sales et al. 2009; Dutton et al. 2010, 
Guo et al. 2010). This could systematically bias the derived rotational properties of the disk if
the mass fraction of the extended gaseous component is large.

\section{Galactic Disk Model}

We adopt the model by Mo, Mao \& White (1998) of an exponential disk,
embedded in a NFW (Navarro et al. 1997) dark matter halo with density distribution

\begin{equation}
\rho_{DM}(r) = \frac{4 \rho_c}{(r/r_s)(1+r/r_s)^2}
\end{equation}

\noindent where $r_s$ is the halo scale radius and $\rho_c$ is 
the dark matter density at $r_s$. The scale radius
is related to the virial radius $r_{200}$ via $r_s=r_{200}/c$
where $c$ is the halo concentration parameter. 
The dark halo rotation curve corresponding to equation 12 is

\begin{equation}
v_{DM}^2(r)=V_{200}^2 \left( \frac{r_{200}}{r} \right) \frac{ln(1+r/r_s)-(r/r_s)/(1+r/r_s)}{ln(1+c)-c/(1+c)}.
\end{equation}

\noindent High-resolution numerical
Cold dark matter (CDM) simulations (e.g. Zhao et al. 2009) show that c depends strongly on cosmological redshift. While c 
decreases with halo virial mass $M_{200}$ at low redshifts, the concentration is
roughly constant with $c \approx 4$ and independent of halo mass at $z \approx 2$.
The halo virial parameters $r_{200}$ and $M_{200}$ are related to each other through the
virial velocity $V_{200}$ (Mo, Mao \& White 1998)

\begin{eqnarray}
r_{200}(z) = \frac{V_{200}(z)}{10 H(z)}, \ \ \ \ \ \
M_{200}(z) = \frac{V_{200}^3(z)}{10 G H(z)}
\end{eqnarray}

\noindent $H$ is the Hubble parameter that depends on cosmological redshift z:

\begin{equation}
H = H_0 \left[ \Omega_{\Lambda} + (1-\Omega_{\Lambda} -
\Omega_M)(1+z)^2 + \Omega_M(1+z)^3 \right]^{1/2} .
\end{equation}

\noindent We adopt a standard $\Lambda$CDM cosmology with
$H_0$ = 73 km/s/Mpc, $\Omega_M$ = 0.238 and $\Omega_{\Lambda} = 0.762$.

The galactic disk is assumed to follow an exponential surface density profile (equation 10).
Its surface density at $r=0$, $\Sigma_0$, is determined by the total disk mass 
$M_d = m_d \times M_{200}$ with $m_d$ the disk mass fraction of the galaxy 

\begin{equation}
\Sigma_0 = m_d \frac{M_{200}}{2 \pi r_{d}^2} .
\end{equation}

\noindent The circular velocity curve of an exponential disk is (Freeman 1970)

\begin{equation}
v_{disk}^2(r) = 4 \pi G \Sigma_0 \ r_{d} y^2 [I_0(y) K_0(y) - I_1(y) K_1(y)]
\end{equation}

\noindent with $y = r/(2 r_{d})$ and the $I_n$ and $K_n$ denoting the modified Bessel functions
(Binney \& Tremaine 2008).

We will neglect a bulge because we are interested in the outer disk parts
where the bulge contribution to the rotation curve is in general negligible. In addition,
several of the best resolved SINS galaxies show no evidence for the presence of a significant
bulge component (Genzel et al. 2008). In this case and 
including adiabatic contraction (Blumenthal et al. 1986; Jesseit et al. 2002) of the dark halo, the zero-pressure
rotation curve is determined from the implicit equation

\begin{eqnarray}
v_0^2(r) & = & v_{disk}^2(r) + v_{DM}^2(r') \\
r' & = & r \left[ 1+\frac{r \times v_{disk}^2(r)}{r' \times v_{DM}^2(r')} \right] .
\end{eqnarray}

\noindent Given $v_0(r)$ the pressure-corrected rotation curve $v_{rot}(r)$ can be
calculated from equation 2 or 11. This is easily done in an iterative process.
One first determines the disk rotation curve, neglecting pressure as
discussed in MMW. Adopting a value of $v_{\max}/\sigma$, its maximum rotational velocity 
provides a first guess for $\sigma$ which leads to a revised rotation curve and a corresponding
new value of v$_{max}$ and $\sigma$. We find that this procedure converges quickly after 10-20 iterations.
In the following we will call $v_{rot}(r)$ the pressure corrected MMW rotation curve.
It can be compared directly with observations (section 4.2). In addition we can calculate
the total disk angular momentum $J_{d} = 2 \pi \int \Sigma v_{rot} rdr$ that will be used
in the next section in order to derive the disk spin parameter $\lambda_d$.

\section{Angular Momentum and Baryon Content of High-Z Galaxies}

Figure 1 shows the half-light radii $r_{1/2}$ of the SINS high-redshift galaxies
versus their maximum rotational velocity $v_{max}$. The data points and
potential uncertainties are discussed in F\"orster-Schreiber et al. (2009),
Cresci et al. (2009) and Law et al. (2009). The errors in $r_{1/2}$ and $v_{max}$
are of order 1-2 kpc and 20-30 km/s, respectively.  We take $r_{1/2}$ 
instead of the exponential disk scale length as it is independent
of any assumption about the light profile. $v_{max}$ is in general a good approximation
of the disk's rotational velocities outside of $r_{1/2}$.
The SINS galaxies segregate strongly into two distinct classes at a critical value
of $v_{max}/\sigma \approx 3$.
We therefore empirically define {\it dispersion-dominated} galaxies (open triangles and stars
in figure 1) as objects with $v_{max}/\sigma \leq 3$.
For these galaxies turbulent pressure gradients have to be included in the interpretation
of the rotation curve.
In contrast, for {\it rotation-dominated} galaxies (filled triangles), defined 
by $v_{max}/\sigma > 3$, pressure effects are small.  Note that $\sigma$ refers to the 
intrinsic velocity dispersion in the disk, not to the observed line-of-sight
or galaxy-integrated dispersion. 
Figure 1 shows that most of the dispersion-dominated galaxies have radii of order 1-3 kpc while the radii of
rotationally dominated galaxies are on average a factor of 2-3 larger. In addition,
the dispersion-dominated systems have rotational velocities of order 100 km/s
while rotation-dominated galaxies rotate with 250 km/s.

The specific angular momentum of a dark halo
is usually specified by the dimensionless spin parameter
(Bullock et al. 2001; Burkert 2009)

\begin{equation}
\lambda = \frac{J_{200}}{\sqrt{2} M_{200} V_{200} r_{200}}
\end{equation}

\noindent where $J_{200}$ is the total angular momentum of the halo. $\lambda$
follows a log-normal distribution with a median of $\lambda = 0.035$ 
and a dispersion of 0.55 (Bullock et al. 2001; Hetznecker \& Burkert 2006).
Cosmological simulations indicate
that in the early phases of protogalactic collapse the gas and dark matter are well mixed,
acquiring similar specific angular momenta (Peebles 1969; Fall \& Efstathiou 1980; White 1984). 
If angular momentum were
conserved during gas infall and all the gas would settle into the disk, 
the resulting disk's specific angular momentum $J_{d}/M_{d}$ would
be similar to the specific angular momentum of the surrounding dark halo.
We cannot measure $\lambda$ directly, but instead can estimate the disk spin parameter, defined as 

\begin{equation}
\lambda_d = \frac{J_{d}}{ \sqrt{2} M_{d} V_{200} r_{200}} = \lambda \frac{j_d}{m_d}
\end{equation}

\noindent with j$_d \equiv $J$_d$/J$_{200}$. If the specific angular momentum of the
infalling gas and the resulting disk is equal to the dark halo's specific angular momentum it
follows that $\lambda_d = \lambda$.

Numerical simulations of galaxy formation find
substantial angular momentum loss of the infalling gas component 
(for a review see e.g. Burkert, 2009). Its origin is not completely clear up to now
and might be attributed to numerical problems or missing physics (for a review see
e.g. Mayer et al. 2008). If the numerical calculations are however correct,
galactic disks should have specific angular momenta and values of $\lambda_d$ 
that are smaller than those of dark matter halos, i.e. on average $\lambda_d \leq 0.035$.

In the following we will investigate disk properties for redshift z=2.2 galaxies with
dark halo concentrations of c=4 and for given values of $\lambda_d$, $m_d$ and
$v_{\max}/\sigma$. We start with 
a first guess of the dark matter virial mass (typically $M_{200}=10^{12}$ M$_{\odot}$). 
Given $m_d$ and by this M$_{d}$ and assuming a disk scale radius r$_d$  and a
$v_{\max}/\sigma$ the procedure discussed in the previous section
gives the corresponding disk rotation curve and by this the corresponding $\lambda_d$.
In an additional iterative step r$_d$ is now varied till the required value of
$\lambda_d$ is achieved.

\begin{figure}[!ht]
\epsscale{1.}
\plotone{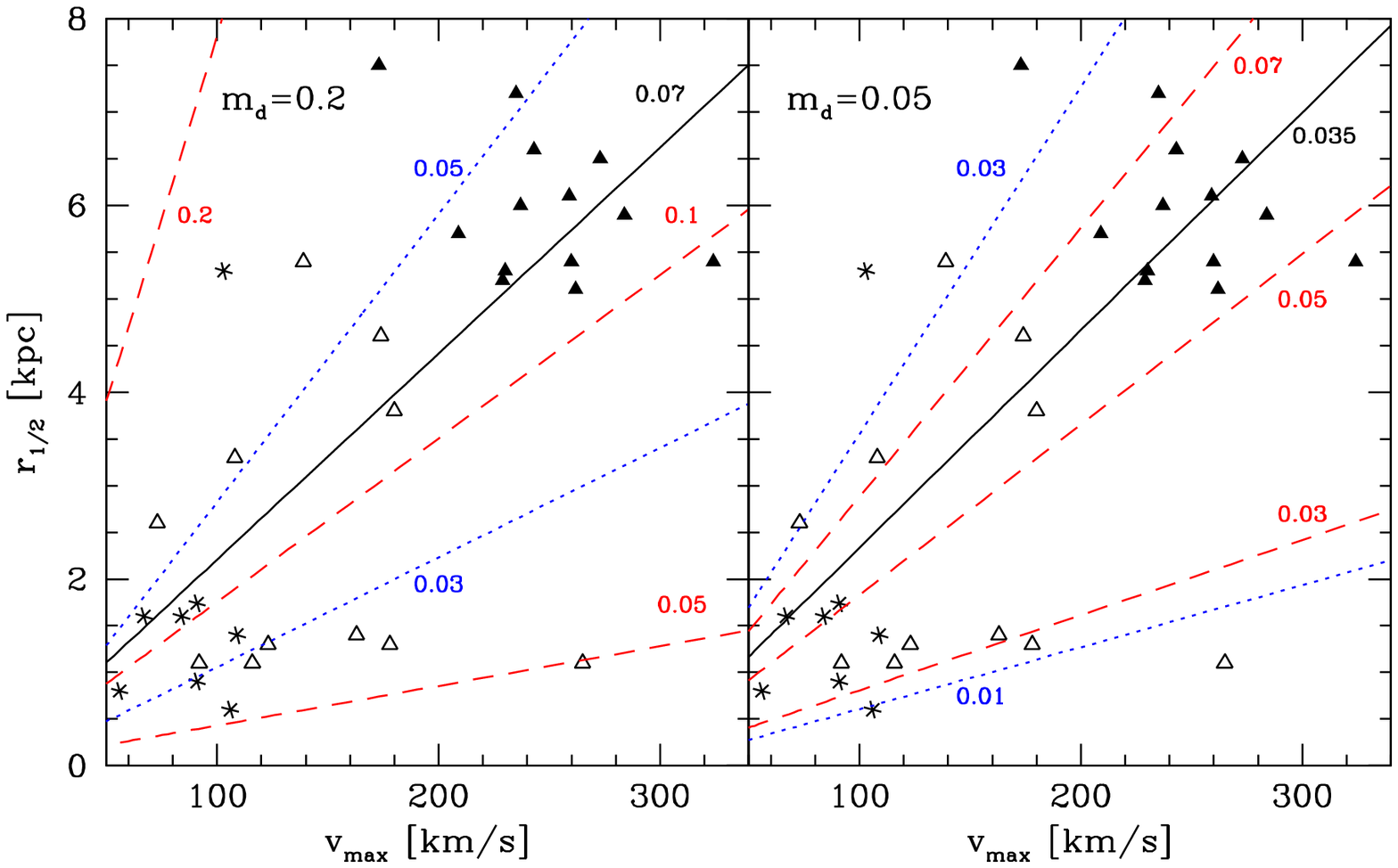}
\caption{
\label{fig1} 
The left and right panels show the disk half-light radii $r_{1/2}$ versus the 
maximum rotational velocities $v_{max}$ of models with disk mass fractions of
m$_d = 0.2$ and m$_d = 0.05$, respectively and compare them with the SINS high-redshift disk sample. 
Open triangles correspond to dispersion-dominated galaxies, filled triangles 
to rotation-dominated objects. Stars show extremely dispersion-dominated systems with
v$_{max}/\sigma \leq 1.5$.
Red dashed lines show the theoretically predicted correlation
between $r_{1/2}$ and $v_{max}$ if pressure effects are neglected for various values
of the disk spin parameter $\lambda_d$ (red labels) 
Blue dotted lines show MMW models including the effect of a pressure gradient and adopting $v_{max}/\sigma = 2$.
The black solid lines represent
rotation-dominated galaxies with $v_{max}/\sigma = 5$, neglecting
adiabatic dark halo contraction.}
\end{figure}

Red dashed lines in figure 1 show the standard MMW model predictions without correcting for pressure effects for a given 
disk spin parameter $\lambda_d$ (red labels), adopting 
a high disk mass fraction, equal to the cosmic baryon fraction (m$_d$ = M$_d$/M$_{200}$ = 0.2) 
in the left panel and a low value of
m$_d=0.05$ in the right panel. $r_{1/2}$ is determined from the known disk scale length: 
$r_{1/2} = 1.68 \times r_d$.  Stars and open triangles correspond
to dispersion-dominated systems, filled triangles to rotation-dominated galaxies. 
Here we assume that the observed half-light radius,
traced by H$\alpha$, is similar to the half-mass radius of the disk. For m$_d = 0.2$ especially the rotation-dominated
galaxies require very large spin parameters, $\lambda_d \approx 0.1-0.2$ 
which are not in agreement with the theoretical expectations of $\lambda_d \leq 0.035$.
Adopting m$_d = 0.05$ improves the situation considerably. The red dashed lines in the right panel
of figure 1 show that, now, MMW models with 
$\lambda_d \approx 0.03 - 0.07$ fit even the fast rotators, consistent with the upper half of
the dark halo $\lambda$ distribution.

\begin{figure}[!ht]
\epsscale{1.}
\plotone{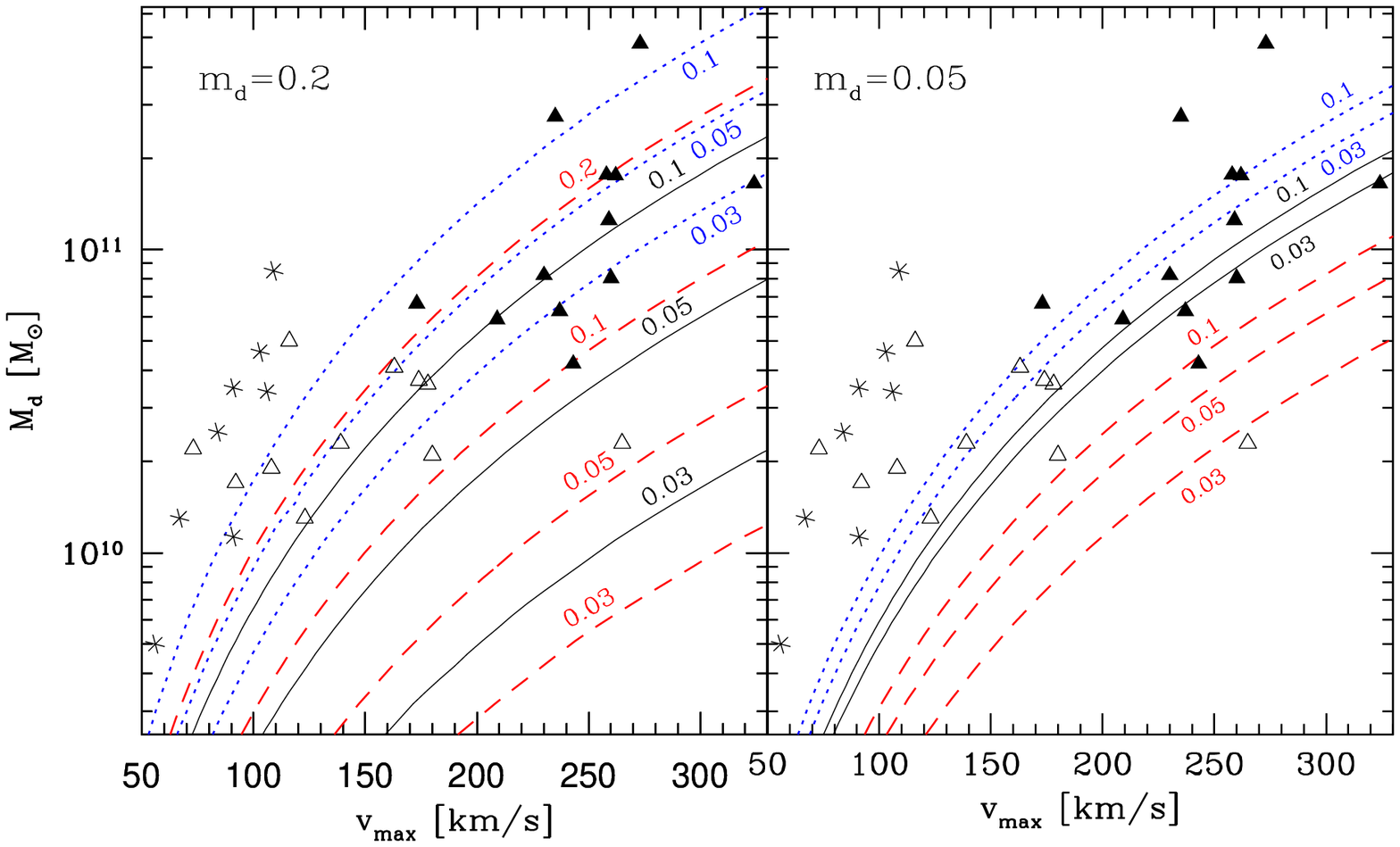}
\caption{
 \label{fig2} 
Open and filled triangles show the observationally inferred disk masses versus
the maximum velocity of dispersion-dominated and rotation-dominated high-redshift SINS galaxies,
respectively. The stars show extremely dispersion-dominated galaxies with $v_{max}/\sigma \leq 1.5$.
The red dashed curves show the predictions of the
standard MMW model, adopting a disk mass fraction of
m$_d = 0.2$ in the left panel and m$_d = 0.05$ in the right panel.
Red labels indicate the corresponding disk $\lambda_d$ parameters.
The blue dotted lines show the situation if turbulent pressure is taken into account,
adopting  $v_{max}/\sigma =2$.  Black, solid curves correspond
to MMW models with $v_{max}/\sigma =5$, neglecting adiabatic contraction of the dark halo.}
\end{figure}

The observed baryonic disk masses provide an additional constraint for theoretical models.
The symbols in figure 2 show the sum of the 
stellar mass (from spectral energy distribution analysis) and gas mass (from an application of the Kennicutt-Schmidt star formation 
relation) of our galaxy sample, plotted as function of $v_{max}$. McGaugh et al. (2000) and
McGaugh (2005) find a remarkably tight correlation between baryonic mass and the circular
velocity v$_{circ}$ at a radius where the rotation curve becomes flat (baryonic Tully Fisher relation):
M$_{d}$/M$_{\odot}$ = 50 $\times (v_{circ}/(km/s))^4$. Interestingly, the high-redshift data shown in figure 2 
deviates strongly from this correlation.
The red dashed curves show the expected correlation between disk mass and $v_{max}$ for different
values of $\lambda_d$ and for large (m$_d = 0.2$, left panel) and small (m$_d = 0.05$, right panel) disk mass fractions 
according to the MMW model, neglecting pressure effects.
According to figure 1, spin parameters of $\lambda_d \approx 0.05$ are required
for m$_d = 0.05$. The right panel of figure 2 however demonstrates that these values are not
consistent with the observed disk masses. On the other hand, the left panels of figure 1 and figure 2 
show that large disk mass fractions of m$_d=0.2$ and spin parameters of  $\lambda_d \approx 0.1-0.2$
are consistent with the observations. While disk mass fractions close to the cosmic baryon fraction
are reasonable for these young galaxies where galactic winds might not yet have removed measurable
amounts of gas, the required large spin parameters are not consistent with the expectation
of $\lambda_d \leq \lambda$, as discussed earlier.

\subsection{Dispersion-Dominated Galaxies and the Importance of Pressure Effects}

In section 2 we demonstrated that pressure gradients can significantly affect
the rotation curves of galaxies when the ratio or rotational velocity to velocity dispersion,
characterized e.g. by $v_{max}/\sigma$, is sufficiently small. In order to compare
the theoretical model with observations we will use the maximum velocity instead of an 
average velocity, e.g. $v_{2.2}$, measured
at 2.15 disk scale lengths. In general this could be dangerous as 
$v_{max}$ could occur anywhere in the disk at radii that are not observed. 
However, as demonstrated by figure 3 for the case of BzK-15504, we find that the
rotation curves in general show an extended flat plateau with the maximum at
1-2 disk scale lengths which is in the observed radius regime. In 
this case, $v_{max}$ is a good approximation of the typical velocity within the flat part of
the rotation curve.

The blue dotted lines in figure 1 show the correlation between $r_{1/2}$ and $v_{max}$ for a
MMW model with pressure correction (equation 11), assuming $v_{max}/\sigma =2$ 
which is consistent with the dispersion-dominated galaxy sample.
The rotation curves are calculated by adopting an exponential
disk with a given half-light radius r$_{1/2}=1.68 \times $r$_d$ (equation 10) and then calculating 
iteratively the corresponding rotation curve as discussed in section 3. Note that $v_{max}$ is now
the maximum of the pressure corrected rotation curve which is smaller than the
value, neglecting pressure effects. With pressure correction most of the pressure-dominated 
galaxies lie in the regime $0.02 \leq \lambda_d \leq 0.05$ for m$_d = 0.2$ and 
$0.01 \leq \lambda_d \leq 0.03$ for m$_d = 0.05$
which is in good agreement with theoretical expectations. 

A significant pressure contribution reduces significantly v$_{max}$ for a given disk
mass M$_d$. The blue dotted lines in figure
2 demonstrate this effect. Like in figure 1, they correspond to disks with $v_{max}/\sigma =2$.
Now, the observed masses of pressure-supported SINS galaxies, represented
by open triangles, are consistent 
with spin parameters $\lambda \approx 0.03 - 0.1$, independent of the adopted value of m$_d$.

The stars in figure 2 show galaxies with $v_{max}/\sigma \leq 1.5$. These galaxies are characterised by
exceptionally low values of v$_{max} \leq 100$ km/s despite their large masses of 
M$_d \approx 10^{10}$ M$_{\odot}$ as expected from equation (11) due to the large pressure contribution.

\subsection{Rotation-Dominated Galaxies and Adiabatic Halo Contraction}

The rotation-dominated SINS sample is characterized by average values of  $v_{max}/\sigma \approx 5$.
The problem of unusually high spin parameters and baryon fractions therefore cannot
be solved by consideration of pressure gradients. A typical representative of this group is
BzK-15504 which has been observed with high angular resolution (Genzel et al. 2006). BzK-15504
is an actively star-forming $z =2.4$ galaxy with a stellar disk mass of 
$10.9^{+2.7}_{-0.1} \times 10^{10}$ M$_{\odot}$, adopting a Chabrier (2003) initial mass function
and a gas mass that varies between
$2.8 \pm 0.6 \times 10^{10}$ M$_{\odot}$ and $4.9 \pm 1.3 \times 10^{10}$ M$_{\odot}$, depending
on the extinction correction applied to the H$\alpha$ line luminosity (F\"orster-Schreiber et al. 2009).
The total disk mass is then $M_d \approx 1.4 \times 10^{11} M_{\odot}$. The radial H$\alpha$ gas 
surface brightness and the rest-frame optical stellar light distributions are both
consistent with an exponential profile with scale length of 4.1 kpc.

The observed line width indicates irregular gas motions of $\sigma \approx 45 \pm 20$ km/s
that are constant throughout the disk outside of the central 3 kpc where a bar and
an active galactic nucleus (AGN) strongly affect the gas kinematics. The maximum rotational velocity is $v_{max}=258 \pm 25$ km/s, so
that $v_{max}/\sigma$ = 5.7.
The dark matter virial mass is not well constrained. A minimum value can be inferred if one
adopts a disk mass fraction, close to the cosmic baryon fraction $f_b = 0.2$:
$M_{200} \geq M_d/f_b \approx 8 \times 10^{11} M_{\odot}$.

\begin{figure}[!ht]
\epsscale{1.}
\plotone{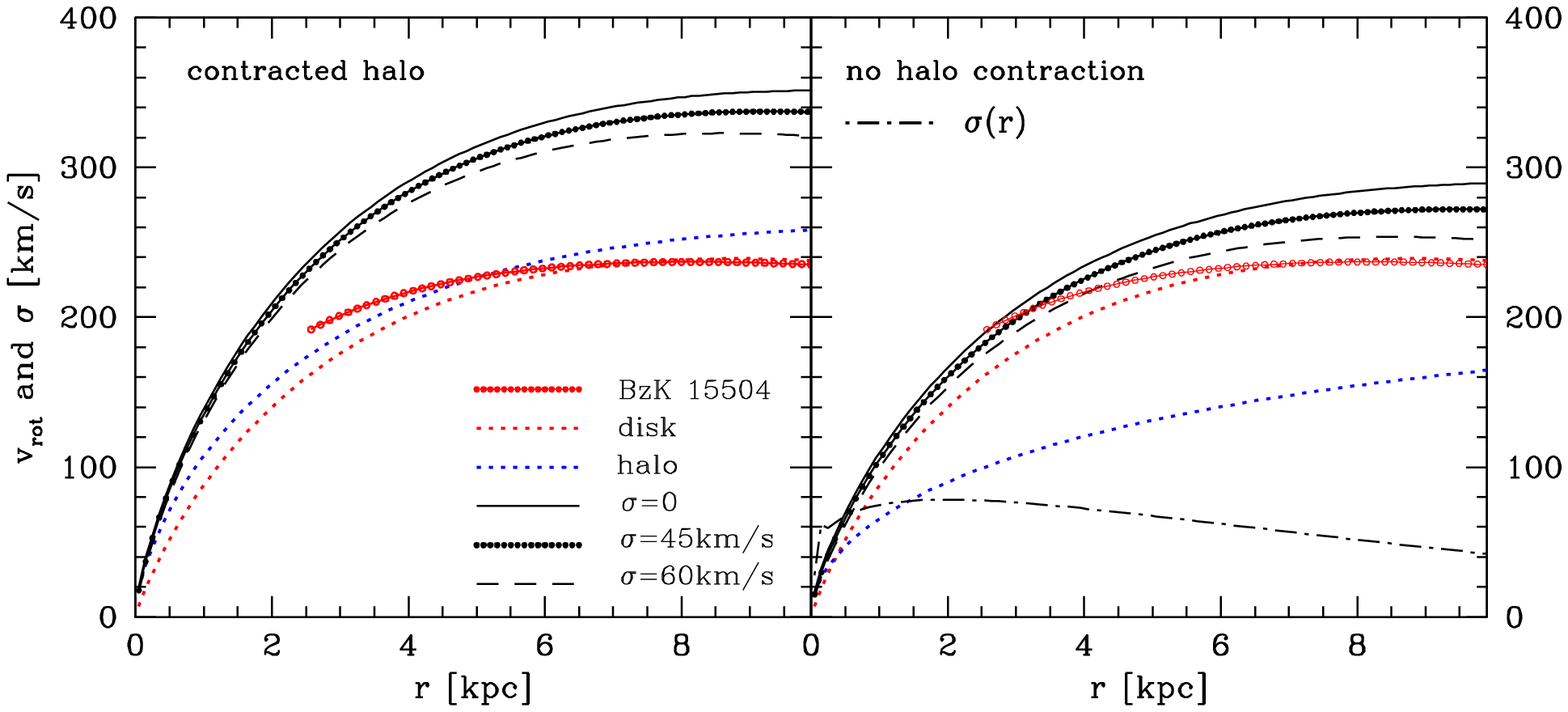}
\caption{
 \label{fig3} 
Open red circles in both panels show the inclination and resolution corrected 
intrinsic rotation curve of BzK-15504, inferred
from fitting the observed two-dimensional distribution of H$\alpha$ velocities
(Genzel et al. 2006). We focus on the rotational properties outside of 3 kpc as the
inner regions are affected by the central AGN and bar. The dotted red and blue
lines in the left panel show the theoretically expected contribution of the disk and dark halo
component, respectively, adopting a cosmic baryon fraction m$_d$ = f$_b$ = 0.2 and an 
adiabatically contracted dark halo with concentraction c=4.
The combination of both curves leads to the zero-pressure total rotation curve 
(upper black line) that exceeds the observed
maximum rotational velocity by more than 100 km/s. The black points and the dashed black curve
correspond to the pressure corrected rotation curve, including pressure effects
with a gas velocity dispersion of 45 km/s and 65 km/s, respectively.
The right panel shows the situation without adiabatic dark halo contraction.
Symbols and lines are the same as in the left panel. Now the resulting rotation curve 
is in much better agreement with the observations.
The lower dot-dashed black curve shows
the theoretically predicted velocity dispersion profile assuming a constant
Toomre stability parameter of Q=1.}
\end{figure}

The dotted red and blue curves in the left panel of 
figure 3 show the disk and dark matter rotation curves, respectively,
adopting a MMW model with the above mentioned disk parameters and M$_{200} = 8 \times 10^{11}$ M$_{\odot}$.
Both components are equally important with peak rotational velocities at r $\approx$ 10 kpc
of 250 km/s and 280 km/s. The upper black curve shows the resulting combined rotation curve,
neglecting pressure effects. It peaks at 370 km/s and is clearly inconsistent with
the observations (red open circles) that peak at $\sim$ 260 km/s.
Including  turbulent pressure with $\sigma$ = 45 km/s does not significantly change
the rotation curve (black points). A test calculation, adopting $\sigma$ = 45 km/s shows that increasing M$_{200}$ 
(decreasing m$_d$) makes the problem even worse while decreasing M$_{200}$ would require
that the galxy has a baryon fraction larger that the cosmic value.

Is this disagreement an observational problem? The uncertainty in the measured disk velocity dispersion is
large, of order 20 km/s. However a dispersion of even 65 km/s (dashed black curves in figure 3) 
makes no big difference. The uncertainties in the determination of the rotation curve are 
of order 25 km/s, too small compared with the disagreement of almost 100 km/s.
The only possible solution appears to be a strong reduction of either the baryonic or the dark matter
mass within the inner 10 kpc. A test calculation shows that the
baryonic disk mass would have to be reduced by a factor of 2 to 
$\sim 7 \times 10^{10} M_{\odot}$ for a pressure corrected MMW model with
velocity dispersion of 65 km/s to fit the observations. 
The estimates of the stellar and gas masses are subject to many uncertainties
and systematics. Spectral energy distribution fitting for BzK-15504 using the Maraston (2005) models
yields stellar masses of $M_* = 9.4 (+2.6/-0.2) \times 10^{10} M_{\odot}$ and gas masses of
$M_{gas} = 3.1 (+0.6/-0.6) \times 10^{10} M_{\odot}$ using the Schmidt-Kennicutt relation
from Bouch\'{e} et al. (2007) and the same extinction
towards HII regions as towards the stars. The gas masses would
be a factor of 2 higher if the extinction towards HII regions is a factor of 2 higher than
towards stars (Calzetti et al. 2004). The best-fit Bruzual \& Charlot
(2003)  model gives $M_* = 10.9 (+2.7/-0.1) \times 10^{10} M_{\odot}$ and
$M_{gas} = 2.8 (+0.5/-0.6) \times 10^{10} M_{\odot}$. Unless the stellar initial mass function
is bottom-light or top-heavy it therefore seems difficult to decrease substantially
the baryonic disk mass of the galaxy.

Another possibility is a smaller dark halo mass in the disk
region. One critical assumption that enters the MMW model is that the dark halo
reacts to the formation of the galactic disk by contracting adiabatically. The right panel 
of figure 3 shows the situation for a MMW model, neglecting dark halo contraction. 
A comparison with the left panel demonstrates the strong effect of adiabatic contraction. 
Although the dark matter virial parameters
in both cases are the same, without adiabatic contraction the contribution of the dark halo 
(blue dashed line) within the 
disk region is small, leading to much better agreement of the model with the observations,
especially if we take into account a turbulent pressure, corresponding to 
a velocity dispersion of $\sigma$ = 65 km/s (upper dashed black line),
that is still within the observed uncertainties and expected if the disk velocity
dispersion tensor would be anisotropic (Aumer et al. 2010).

The problem discussed for BzK 15504 exists for all rotation-dominated galaxies that are
represented in the figures 1 and 2 by black filled triangles. 
Due to their large values of $v_{max}/\sigma \approx 5$ the effect of pressure gradients is small.
The galaxies are therefore represented well by the dashed red lines in both figures 
which indicate that even for large disk spin parameters $\lambda_d = 0.1$ the maximum rotational
velocity exceeds the observations for given M$_d$.
The solid black lines in both figures show the strong effect of neglecting
adiabatic dark halo contraction. According to figure 1, the observed disk radi require
$\lambda_d \approx 0.07$ for m$_d = 0.2$ and $\lambda_d = 0.035$ for 
m$_d = 0.05$. Figure 2 shows that these spin values are also consistent
with the observed disk masses.

\section{Origin of Gas Turbulence in High-Redshift Disk Galaxies}

The MMW models also provide insight into the origin of the observed gas turbulence.
Let us propose that the main driver of clumpiness and turbulence
in gas-rich high-redshift disks is gravitational disk instability.
Then we expect gas-rich disks to stay close to 
the gravitational stability line because of the following reason. A disk that is kinematically too cold
with small velocity dispersions is highly gravitationally unstable. Gravitational instabilites
generate density and velocity irregularities that drive turbulence and heat
the system kinematically. As a result, the gas
velocity dispersion increases till it approaches the stability limit where kinetic driving by gravitational 
instabilities saturates. A disk with even higher velocity dispersions would be stable.
Here the turbulent energy would
dissipate efficiently and the velocity dispersion would decrease again until it crosses
the critical velocity dispersion limit where gravitational instabilities become efficient again in 
driving turbulent motions. In summary,
galactic disks should settle close to the gravitational stability line that is determined
by the Toomre criterion (Toomre 1964; Wang \& Silk 1994)

\begin{equation}
Q \equiv \frac{\kappa}{\pi G} \left( \frac{\Sigma_g}{\sigma_g} + \frac{\Sigma_*}{\sigma_*} 
  \right)^{-1} \leq Q_c.
\end{equation}

\noindent $\kappa$ is the epicyclic frequency that is related to the local angular circular
velocity $\Omega$ at radius r through $\kappa^2 = r d \Omega^2/dr+4 \Omega^2$
and $Q_c$ is the critical value which is of order unity
(Goldreich \& Lynden-Bell 1965; Dekel et al. 2009b).
$\Sigma_*$ and $\sigma_*$ are the stellar surface density and velocity dispersion, respectively.
As a test, let us focus again on BzK 15504.
As most of the stars in BzK 15504 are likely to have formed during the presently observed
star burst we can assume that the stellar velocity dispersion is similar to the observed turbulent
gas velocity, i.e. $\sigma_* \approx \sigma_g$. In addition, the observations show that
both components have similar exponential disk scale lengths.
If the disk is close to the instability line, its turbulent gas velocity dispersion at any point r 
is then

\begin{equation}
\sigma = \frac{\pi G \Sigma}{\kappa}
\end{equation}

\noindent where $\Sigma (r)$ is the local baryonic (gas+stars) disk surface density.
The lower black dot-dashed curve in the right panel of figure 3 shows the predicted gas velocity dispersion 
for BzK-15504.  It is indeed almost independent of radius and within the uncertainty
in agreement with the observed, radially constant value of 45 $\pm$ 25 km/s.
We thus conclude that BzK-15504 is a marginally unstable star-forming disk
(Genzel et al. 2006), driven by gravitational instabilities.

We calculated the velocity dispersion profile for all SINS galaxies with $v_{max}/\sigma \geq 2$
using a pressure corrected MMW model with a dark halo concentration c=4
and neglecting adiabatic halo contraction. The disk mass
was taken from the observed stellar and gas masses. The dark halo mass and by this m$_d$
was constrained by fitting
the observed maximum velocity of the galaxies. In all cases the theoretically derived velocity dispersion $\sigma_{theo}$,
adopting equation 23, is almost constant within 1 and 2 disk scale radii. Figure 4 compares the
average value of $\sigma_{theo}$ within 1 and 2 r$_d$ 
with the observed velocity dispersion $\sigma_{obs}$. Dispersion-dominated (open triangles)
and rotation-dominated (filled triangles) systems have similar gas velocity dispersions, indicating that the
difference in  $v_{max}/\sigma$ is due to a difference in rotational velocities and not a result of differences
in the turbulent gas velocity. Despite the large uncertainties, the
theoretical and observed velocity dispersions agree well, strengthening the suggestion that gravitational
instabilities are the major driver of turbulence in high-redshift star-forming galaxies.

\begin{figure}[!ht]
\epsscale{1.}
\plotone{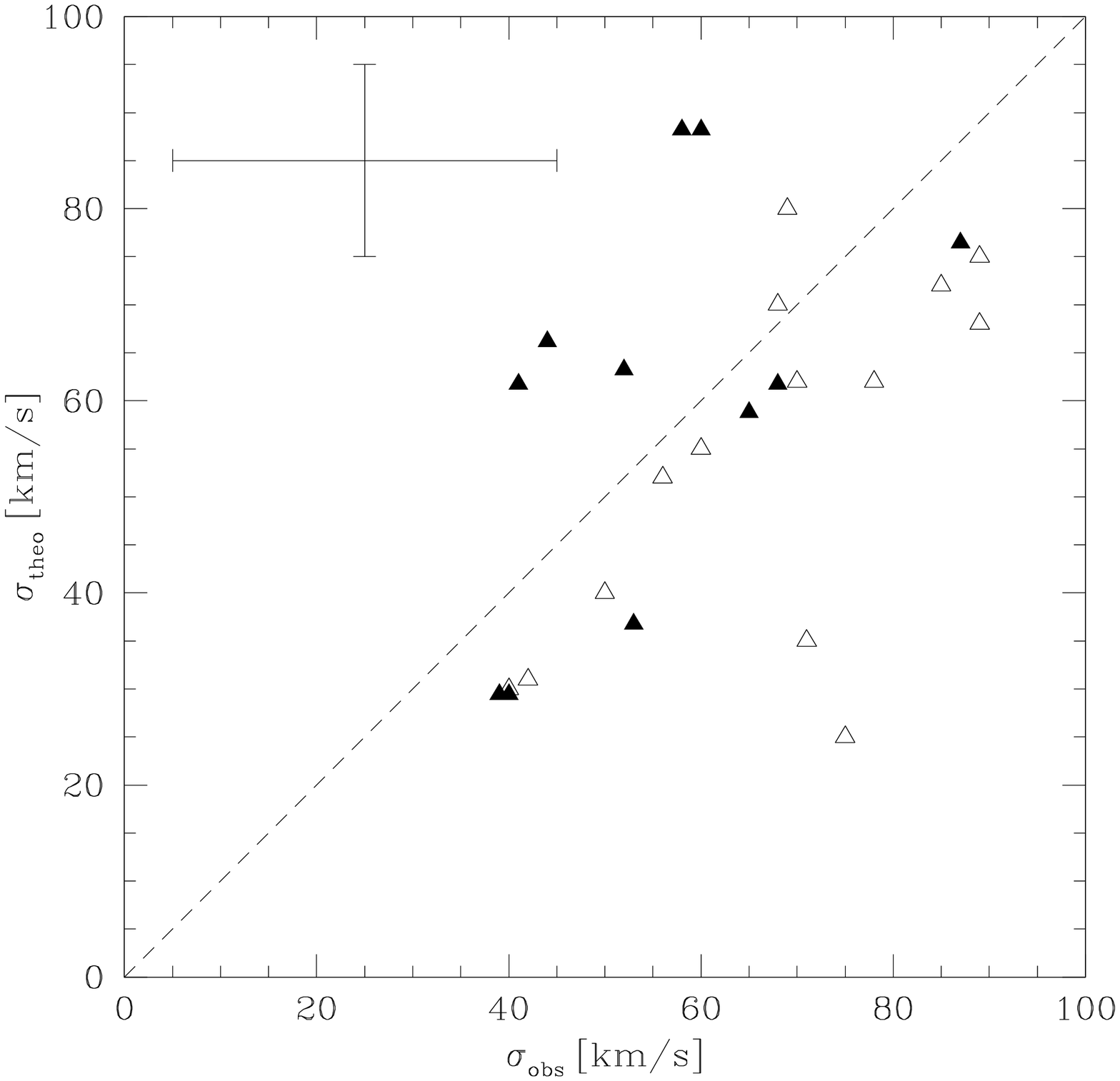}
\caption{
 \label{fig4} The observed velocity dispersion $\sigma_{obs}$ of the SINS high-redshift galaxy
sample is compared with theoretical expectations $\sigma_{theo}$, adopting a pressure-corrected MMW model
without adiabatic halo contraction. $\sigma_{theo}$ was calculated by averaging
the velocity dispersion profile between one and two disk scale lengths.
Open and filled triangles correspond to dispersion and rotation-dominated systems, 
respectively. The error bar in the upper left corner indicates the observational uncertainties.}
\end{figure}

\section{Summary and Discussion}

We have shown that pressure gradients in turbulent galactic disks can
significantly affect their rotation curves. This effect is well-known for thick stellar disks,
leading e.g. to an asymmetric drift of kinematically hot stellar populations in
the Galaxy (Binney \& Tremaine 2008). A similar effect is found in 
models of dust growth in protoplanetary disks where dust particles
on ballistic orbits rotate faster than the disk gas which rotates sub-Keplerian due 
to pressure gradients, leading to fatal dust migration into the central star
(e.g. Takeuchi \& Artymowicz 2001). 

We analysed the SINS sample of dispersion-dominated high-redshift star-forming galaxies.
Including pressure effects and adopting an exponential gas disk with scale length similar
to the stellar disk, the models can explain the properties
of the dispersion-dominated SINS galaxies very well, with disk spin parameters of
$\lambda_d \approx 0.025 - 0.05$ and disk mass fractions of $m_d \approx 0.05-0.2$.
The strong pressure effect on the structure of dispersion-dominated galaxies
depends critically on the assumption of a gas pressure gradient in their disks.
The best resolved SINS galaxies
indicate a radially constant turbulent velocity dispersion
and an exponential decline of the surface density of star-forming gas with scale
length similar to the stellar disk (Bouch\'{e} et al., in preparation).
It is however not clear whether all high-redshift galaxy have such a structure.
More high-resolution observations are required in order to clarify this point.

For rotation-dominated galaxies, defined by $v_{max}/\sigma \geq 3$,
pressure gradients cannot strongly affect the rotational velocities. 
Analysing as a test case the galaxy BzK 15504
we showed that its rotation curve allows no significant
contribution of dark matter within the visible disk region. 
This can be achieved with a standard NFW halo that did
not contract adiabatically in response to the formation of the galactic disk.
The MMW models of the rotation-dominated sample, neglecting adiabatic dark halo contraction, lead
to reasonable values of $\lambda_d \approx 0.035$ if the disk mass fraction is low
(m$_d \approx 0.05$) while in the case of high mass fraction (m$_d \approx 0.2$), the 
spin parameters would again be extrem ($\lambda_d \approx 0.1$).

We assumed that high disk spin parameters of $\lambda_d \geq 0.1$ are unreasonable. 
Scenarios have however been constructed that could
lead to disks with higher $\lambda_d$ values then their dark matter halos.
One of the most promising suggestions
are selective galactic outflows of especially low-angular momentum gas from galactic centers
(e.g. Maller \& Dekel 2002; Dutton \& van den Bosch 2009). 

Sales et al. (09) compared MMW model predictions with the SINS high-redshift
disk galaxies and found a good agreement in the rotational velocity versus disk 
scale length plane (similar to Fig. 1) for reasonable spin parameters $\lambda_d \approx 0.04 - 0.06$.
They assumed values of $m_d = 0.05$ and included dark halo contraction. Unfortunately, Sales et al.
do not compare their predicted disk masses with observations. In addition,
they argue that the observed gas disk radii are a factor $\kappa_r = 1.8$ larger than
the corresponding stellar disk. They then multiply the disk scale radii
resulting from their MMW model for a given $\lambda_d$ by this factor in order to compare their model with
observations. This implicitly assumes that the
gas component has a negligible mass as otherwise it would affect $\lambda_d$. 
The assumption of a low gas-to-star fraction is not supported by the SINS observations which typically
indicate disk gas fraction of order 30-60\% (Tacconi et al. 2010). In addition,
for those cases where stellar data is also available Cresci et al. (2009) find similar scale
radii for the stars and gas with $\kappa_r \approx 1$, justifying our assumption.
Still, more complex multi-component MMW models might be interesting in order to 
better understand the physical properties 
of systems with stellar and gaseous disks that are characterised by different scale radii.

The problem that adiabatic halo contraction leads to compact galactic disks
with scaling relations that are not in agreement with observations has been
discussed for low-redshift galaxies e.g. by 
Dutton et al. (2007, with references therein) who advocate a model in which the dark halo
actually expands rather than contracting. We find that this effect is important also for high-redshift galaxies.
Several solutions are currently being discussed.
Gnedin et al. (2004) and Gustafsson et al. (2006) argue that the circular orbit
adiabatic contraction model (Barnes \& White 1984) considerably overestimates the amount of dark matter contraction.
The numerical simulations of Jesseit et al. (2002) however find good agreement with the analytical expression.
The dark matter mass fraction in the disk region could also be reduced if one assumes
cored dark matter halos (Burkert 1995; Salucci \& Burkert 2000), resulting e.g. from 
dark matter annihilation, dark matter particle scattering or dynamical interaction. Halo
expansion could be triggered by
dynamical interaction with massive subclumps or molecular clouds in the disk
(El-Zant et al. 2001; Dutton et al. 2007, Mashchenko et al. 2006; Johansson et al. 2009, Abadi et al. 2010;
Jardel \& Sellwood 2009) or with bars (e.g. Weinberg \& Katz 2002;
Sellwood 2008).  Rapid outflows of gas would also lead to halo expansion
(Navarro, Eke \& Frenk 1996; Gnedin \& Zhao 2002; Read \& Gilmore 2005).

Whatever the origin, our result demonstrates that the problem of inefficient dark halo 
contraction is related to the earliest phases of galaxy formation.
Although not required in order to produce reasonable spin parameters and disk mass fractions, the processes
that suppressed adiabatic halo contraction in rotation-dominated galaxies might also have been active
in dispersion-dominated systems. In this case, most dispersion-dominated systems would be characterized
by even smaller spin parameters of $\lambda_d = 0.01 - 0.03$.

We have analysed the origin of turbulence in high-redshift disk galaxies.
Assuming that the disks are marginally unstable we can explain the observed
velocity dispersion.  This indicates that turbulence 
is driven and regulated by gravitational instabilites, combined with turbulent energy
dissipation. We argued that in this case galactic disks will tend to
stay close to a state of marginal gravitational stability which for gas-rich disks
corresponds to a velocity dispersion of order 40 - 80 km/s. The energetic source of the turbulent driver
is then the potential energy of the disks' gas (Krumholz \& Burkert 2010) which, coupled
with viscous forces releases potential energy by spiraling inwards, 
generating at the end bulge-dominated galaxies as suggested e.g. by 
Elmegreen et al. (2008) and Dekel et al. (2009a,b).
Other energy sources like stellar feedback or accretion energy from infalling 
gas would then play a minor role because if these processes were
dominant the velocity dispersion would likely differ from the value expected for
a marginally unstable disk. Note, that within the framework of this scenario the observed
velocity dispersion is a signature of global gas motions that affect the global disk structure and not just 
the result of local stellar energy feedback, generating HII regions and driving local outflows of ionized gas. 
This conclusion is consistent with the
finding of Elmegreen \& Elmegreen (2006) that the stellar z-scale heights of high-redshift
star-forming galaxies are of order 1 kpc, which translates to a global velocity dispersion of
order 50 km/s. Whether gas-rich galactic disks naturally evolve towards a
state of marginal stability through gravitational driving of turbulence, combined with
turbulent energy dissipation is an interesting question
that should be explored in greater details.

\noindent
{\bf Acknowledgments:}
A.B. is partly supported by a Max-Planck-Fellowship. This work was supported by the DFG Cluster
of Excellence "Origin and Structure of the Universe". The Dark Cosmology Centre is funded by the 
Danish National Research Foundation. We thank A. Dekel, J. Navarro and L. Sales for useful discussions.
We thank the referee for a detailed and constructive report that greatly improved the quality of our paper.


\begin{thebibliography}{99}

\bibitem[Abadi et al. (2010)]{aba10} Abadi, M.G., Navarro, J., Fardal, M., Babul, A. \& Steinmetz, M. 2010, MNRAS, 407, 435

\bibitem[Aumer et al. (2010)]{aum10} Aumer, M., Burkert, A., Johansson, P. \& Genzel, R. 2010, ApJ, 719, 1230

\bibitem[Barnes \& White (1984)]{bar84} Barnes, J. \& White, S.D.M. 1984, MNRAS, 211, 753

\bibitem[Binney \& Tremaine (2008)]{bin08} Binney, J. \& Tremaine, S. 2008, Galactic Dynamics (Princeton, NJ, Princeton University Press, 2008)

\bibitem[Blumenthal et al. (1986)]{blu86} Blumenthal, G.R., Faber, S.M., Flores, R. \& Primack, J.R. 1986, ApJ, 301, 27

\bibitem[Bouch\'{e} et al. (2007)]{bou07} Bouch\'{e}, N. et al. 2007, ApJ, 671, 303

\bibitem[Bournaud, Elmegreen \& Elmegreen (2007)]{bou07} Bournaud, F., Elmegreen, B.G. \& Elmegreen, D.M. 2007, ApJ, 670, 237

\bibitem[Bournaud et al. (2008)]{bou08} Bournaud, F. et al. 2008, A\&A, 486, 741

\bibitem[Bruzual \& Charlot (2003)]{bru03} Bruzual, G. \& Charlot, S. 2003, MNRAS, 344, 1000

\bibitem[Bullock et al. (2001)]{bul01} Bullock, J.S. et al. 2001, ApJ, 555, 240

\bibitem[Burkert (1995)]{bur95} Burkert, A. 1995, ApJ, 447, L25

\bibitem[Burkert (2009)]{bur09} Burkert, A. 2009, in IAU Symp. 254, The Galaxy Disk in Cosmological Context, ed. J. Andersen et al. (Cambridge
Univ. Press), 437

\bibitem[Calzetti et al. (2004)]{cal04} Calzetti, D., Harris, J., Gallagher, J.S., Smith, D.A., Conselice, C.J., Homeier, N. \& Kewley, L. 2004, AJ, 127, 1403

\bibitem[Chabrier (2003)]{cha03} Chabrier, G. 2003, PASP, 115, 763

\bibitem[Courteau et al. (2007)]{cou07} Courteau et al. 2007, ApJ, 671, 203

\bibitem[Cresci et al. (2009)]{cre09} Cresci, G. et al. 2009, ApJ, 697, 115

\bibitem[Daddi et al. (2004a)]{dad04a} Daddi, E. et al. 2004a, ApJ, 600, L127

\bibitem[Daddi et al. (2004b)]{dad04b} Daddi, E., Cimatti, A., Renzini, A., Fontana, A., Mignoli, M., Pozetti, L., Tozzi, P. \& Zamorani, G. 2004b, ApJ,  617, 746

\bibitem[Daddi et al. (2007)]{dad07} Daddi, E. et al. 2007, ApJ, 670, 156

\bibitem[Dekel \& Birnboim (2006)]{dek06} Dekel, A. \& Birnboim, Y. 2006, MNRAS, 368, 2

\bibitem[Dekel et al. (2009a)]{dek09a} Dekel, A. et al. 2009a, Nature, 457, 451

\bibitem[Dekel, Sari \& Ceverino (2009b)]{dek09b} Dekel, A., Sari,, R. \& Ceverino, D. 2009b, ApJ, 703, 785

\bibitem[Dib, Bell \& Burkert (2006)]{dib06} Dib, S., Bell, E. \& Burkert, A. 2006, ApJ, 638, 797

\bibitem[Dutton et al. (2007)]{dut07} Dutton, A.A., van den Bosch, F.C., Dekel, A. \& Courteau, S. 2007, ApJ, 654, 27

\bibitem[Dutton et al. (2009)]{dut09} Dutton, A.A. \& van den Bosch, F.C. 2009, MNRAS, 396, 141

\bibitem[Dutton et al. (2010)]{dut10} Dutton, A.A. et al. 2010, MNRAS in press, astro-ph/1006.3558

\bibitem[Elmegreen \& Elmegreen (2006)]{elm06} Elmegreen, D.M. \& Elmegreen, B.G. 2006, ApJ, 650, 644 

\bibitem[Elmegreen et al. (2007)]{elm07} Elmegreen, D.M., Elmegreen, B.G., Ravindranath, S. \& Cox, D.A. 2007, ApJ, 658, 763

\bibitem[Elmegreen, Bournaud \& Elmegreen (2008)]{elm08} Elmegreen, B.G,, Bournaud, F. \& Elmegreen, D. 2008, ApJ, 688, 67

\bibitem[Elmegreen \& Burkert (2010)]{elm10} Elmegreen, D.M. \& Burkert, A. 2010, ApJ, 712, 294

\bibitem[Elmegreen, Bournaud \& Elmegreen (2008)]{elm08} Elmegreen, B.G., Bournaud, F. \& Elmegreen, D.M. 2008, ApJ, 688, 67 

\bibitem[El-Zant, Shlosman \& Hoffman (2001)]{elz01} El-Zant, A., Shlosman, I. \& Hoffman, Y. 2001, ApJ, 560, 636

\bibitem[Epinat et al. (2009)]{epi09} Epinat et al. 2009, A\&A, 504, 789

\bibitem[Erb et al. (2006a)]{erb06a} Erb, D. K., Steidel, C. C., Shapley, A. E., Pettini, M., Reddy, N. A. \& Adelberger, K. L. 2006a ApJ, 646,107

\bibitem[Erb et al. (2006b)]{erb06b} Erb, D. K., Steidel, C. C., Shapley, A. E., Pettini, M., Reddy, N. A. \& Adelberger, K. L. 2006b, ApJ, 647,128

\bibitem[Fall \& Efsthatiou (2006)]{fal06} Fall, S.M. \& Efsthatiou, G. 1980, MNRAS, 193, 189

\bibitem[F\"orster Schreiber et al. (2006)]{foe06} F\"orster Schreiber, N. M. et al. 2006, ApJ, 645, 1062 

\bibitem[F\"orster Schreiber et al. (2009)]{foe09} F\"orster Schreiber, N. M. et al. 2009, ApJ, 706, 1364

\bibitem[Franx et al. (2003)]{fra03} Franx, M., et al. 2003, ApJ, 587, L79

\bibitem[Freeman et al. (1970)]{fre70} Freeman, K.C. 1970, ApJ, 160, 811

\bibitem[Genel et al. (2008)]{gen08} Genel, S. et al. 2008, ApJ, 688, 789

\bibitem[Genzel et al. (2006)]{gen06} Genzel, R. et al. 2006, Nature, 442, 786

\bibitem[Genzel et al. (2008)]{gen08} Genzel, R. et al. 2008, ApJ, 687, 59

\bibitem[Gnedin \& Zhao (2002)]{gne02} Gnedin, O.Y. \& Zhao, H. 2002, MNRAS, 333, 299

\bibitem[Gnedin et al. (2004)]{gne04} Gnedin, O,Y., Kravtsov, A.V., Klypin, A.A. \& Nagai, D. 2004, ApJ, 616, 16

\bibitem[Goldreich \& Lynden-Bell (1965)]{gol65} Goldreich, P. \& Lynden-Bell, D. 1965, MNRAS, 130, 125

\bibitem[Grazian et al. (2007)]{gra07} Grazian, A. et al. 2007, A\&A, 465, 393

\bibitem[Guo et al. (2010)]{guo10} Guo, Q. et al. 2010, submitted to MNRAS, astro-ph/1006.0106

\bibitem[Gustafsson, Fairbairn \& Sommer-Larsen (2006)]{gus06} Gustafsson, M., Fairbairn, M. \& Sommer-Larsen, J. 2006, Phys. Rev. D., 74, 3522

\bibitem[Hetznecker \& Burkert (2006)]{het06} Hetznecker, H. \& Burkert, A. 2006, MNRAS, 370, 1905

\bibitem[Jardel \& Sellwood (2009)]{jar09} Jardel, J.R \& Sellwood, J.A. 2009, ApJ, 691, 1300

\bibitem[Immeli et al. (2004a)]{imm04a} Immeli, A., Samland, M. Gerhard, O. \& Westera, P. 2004a, A\&A, 413, 547

\bibitem[Immeli et al. (2004b)]{imm04b} Immeli, A., Samland, M. Westera, P. \& Gerhard, O. 2004b, ApJ, 611, 20

\bibitem[Jesseit, Naab \& Burkert (2002)]{jes02} Jesseit, R., Naab, T. \& Burkert, A. 2002, ApJ, 571, 89

\bibitem[Johansson, Naab \& Ostrikea (2009)]{joh09} Johansson, P.H., Naab, T.\& Ostriker, J.P. 2009, ApJ, 697, L38

\bibitem[Khochfar \& Silk (2009)]{kho09} Khochfar, S. \& Silk, J. 2009, ApJ, 700, L21

\bibitem[Kere\u{s} et al. (2005)]{ker05} Kere\u{s} D., Katz, N., Weinberg, D.H. \& Dav\'{e}, R. 2005, MNRAS, 363, 2

\bibitem[Krumholz \& Burkert (2010)]{kru10} Krumholz, M. \& Burkert, A. 2010, ApJ, 724, 895

\bibitem[Law et al. (2007)]{law07} Law, D., Steidel, C.C., Erb, D.K., Larkin, J.E., Pettini, M., Shapley, A.E. \& Wright, S.A. 2007, ApJ, 669, 929

\bibitem[Law et al. (2009)]{law09} Law, D. et al. 2009, ApJ, 697, 2057

\bibitem[Maller \& Dekel (2002)]{mal02} Maller,A.H. \& Dekel, A. 2002, MNRAS, 335, 487

\bibitem[Maraston (2005)]{mar05} Maraston, C. 2005, MNRAS, 362, 799

\bibitem[McGaugh et al. (2000)]{mcg00} McGaugh, S.S., Schombert, J.M., Bothun, G.D. \& de Blok, W.J.G. 2000, ApJ, 533, L99

\bibitem[McGaugh (2005)]{mcg05} McGaugh, S.S. 2005, ApJ, 632, 859

\bibitem[Mo, Mao \& White (1998)]{mo98} Mo, H.J., Mao, S. \& White, S.D.M. 1998, MNRAS, 295, 319

\bibitem[Mo \& Mao (2000)]{mo00} Mo,H.J. \& Mao, S. 2000, MNRAS 318, 163

\bibitem[Mashchenko, Couchman \& Wadsley (2006)]{mas06} Mashchenko, S., Couchman, H.M.P. \& Wadsley, J. 2006, Nature, 442, 7102

\bibitem[Navarro, Eke \& Frenk (1996)]{nav96} Navarro, J.F., Eke, V.R. \& Frenk, C.S. 1996, MNRAS 283, L72

\bibitem[Mayer, Governato \& Kaufmann (2008)]{may08} Mayer, L., Governato, F. \& Kaufmann, T. 2008, ASL, 1, 7

\bibitem[Navarro, Frenk \& White (1997)]{nav97} Navarro, J.F., Frenk, C.S. \& White, S.D.M. 1997, ApJ, 490, 493

\bibitem[Noeske et al. (2007)]{noe07} Noeske, K.G. et al. 2007, ApJ, 660, L43

\bibitem[Ocvirk, Pichon \& Teyssier (2008)]{ocv08} Ocvirk, P., Pichon, C. \& Teyssier, R. 2008, MNRAS,  390, 1326

\bibitem[Peebles (1969)]{pee69} Peebles, P.J.E. 1969, ApJ, 155, 393

\bibitem[P\'{e}rez-Gonzalez et al. (2008)]{per08} P\'{e}rez-Gonzalez, P.G. et al. 2008, ApJ, 675, 234

\bibitem[Read \& Gilmore (2005)]{rea05} Read, J.I. \& Gilmore, G. 2005, MNRAS, 356, 107

\bibitem[Reddy et al. (2005)]{red05} Reddy, N. A., Erb, D. K.,Steidel, C. C., Shapley, A. E., Adelberger, K. L., \& Pettini, M. 2005, ApJ, 633, 748

\bibitem[Rudnick et al. (2006)]{rud06} Rudnick, G. et al. 2006, ApJ, 650, 624

\bibitem[Sales et al. (2009)]{sal09} Sales, L.V. et al. 2009, MNRAS, 399, L64

\bibitem[Salucci \& Burkert (2000)]{sal00} Salucci, P. \& Burkert, A. 2000, ApJ, 537, 9

\bibitem[Sellwood (2008)]{sel08} Sellwood, J.A. 2008, ApJ, 679, 379

\bibitem[Shapiro et al. (2008)]{sha08} Shapiro, K.L. et al. 2008, ApJ, 682, 231

\bibitem[Shapley et al. (2005)]{sha05} Shapley, A.E. et al. 2005, ApJ, 626, 698

\bibitem[Spitzer (1942)]{spi42} Spitzer, L. 1942, ApJ, 95, 329

\bibitem[Stark et al. (2008)]{sta08} Stark et al. 2008, Nature, 455, 775

\bibitem[Steidel et al. (1996)]{ste96} Steidel, C.C., Giavalisco, M., Pettini, M., Dickinson, M. \& Adelberger, K.L. 1996, ApJ, 462, L17

\bibitem[Steidel et al. (2004)]{ste04} Steidel, C.C., Shapley, A.E., Pettini, M., Adelberger, K.L., Erb, D.K., Reddy, N.A., \& Hunt, M.P. 2004, ApJ, 604, 534

\bibitem[Tacconi et al. (2010)]{tac10} Tacconi, L. et al. 2010, Nature, 463, 781

\bibitem[Takeuchi \& Artymowicz (2001)]{tak01} Takeuchi, T. \& Artymowicz, P. 2001, ApJ, 557, 990

\bibitem[Toomre (1964)]{too64} Toomre, A. 1964, ApJ, 139, 1217

\bibitem[van Dokkum et al. (2006)]{van06} van Dokkum, P.G.  et al. 2006, ApJ, 638, L59

\bibitem[Van Starkenburg et al. (2008)]{van08} Van Starkenburg, L., van der Werf, R.P., Franx, M., Labb'{e}, I., Rudnick, G. \& Wuyts, S. 2008, A\&A, 488, 99

\bibitem[Wang \& Silk (1994)]{wan94} Wang, B. \& Silk, J. 1994, ApJ, 427, 759

\bibitem[Weinberg \& Katz (2002)]{wei02} Weinberg, M.D. \& Katz, N. 2002, ApJ, 580, 627

\bibitem[White (1984)]{whi84} White, S.D.M. 1984, MNRAS, 286, 38

\bibitem[Wright et al. (2007)]{wri07} Wright, S.A. et al. 2007, ApJ, 658, 78

\bibitem[Wright et al. (2009)]{wri09} Wright, S.A. et al. 2009, ApJ, 699, 421

\bibitem[Zhao et al. (2009)]{zha09} Zhao, D.H., Jing, Y.P., Mo, H.J. \& Boerner, G. 2009, ApJ, 707, 354

\end{thebibliography}
\end{document}